\documentclass[12pt,tightenlines,eqsecnum,floats,aps,amsmath,amssymb,nofootinbib,prd,showpacs]{revtex4}
\usepackage{subfigure}

\usepackage{setspace}
\usepackage{amsmath,amssymb,amsfonts,amsthm}
\usepackage{graphicx}
\usepackage{enumerate} 
\usepackage{colordvi} 

\def\be{\begin{equation}}
\def\ee{\end{equation}}
\def\ba{\begin{eqnarray}}
\def\ea{\end{eqnarray}}
\def\bi{\begin{itemize}}
\def\ei{\end{itemize}}
\def\bnum{\begin{enumerate}}
\def\enum{\end{enumerate}}

\def\V{\mathcal{V}}
\def\v{\nu}
\def\p{\phi}

\def\H{{\cal H}}
\def\Hkg{\H_{\rm kin}^{\rm grav}}
\def\Hk{\H_{\rm kin}}

\def\Hkm{\H_{\rm kin}^{\rm matt}}

\def\wh{\widehat}

\def\f{\frac}

\def\pp{p}
\def\lp{{\ell}_{\rm Pl}}

\def\lo{\ell_o}
\def\bra{\langle}
\def\ket{\rangle}
\def\sint{\textstyle{\int}}

\def\R{\mathbb{R}}

\def\Z{\mathbb{Z}}

\def\e{\text{e}}
\def\G{A}

\def\dd{\textrm{d}}

\usepackage{enumerate}

\usepackage{colordvi}
\newcounter{mnotecount}[section]

\newcommand{\comment}[1]{}
\begin{document}

\title{Path Integrals and the WKB approximation\\ in Loop Quantum Cosmology}

\author{Abhay Ashtekar} \email{ashtekar@gravity.psu.edu}
\author{Miguel Campiglia} \email{miguel@gravity.psu.edu}
\author{Adam Henderson}\email{henderson@gravity.psu.edu}
 \affiliation{Institute for Gravitation and the Cosmos \& Physics
 Department,
 Penn State, University Park, PA 16802-6300, U.S.A.}

\begin{abstract}

We follow the Feynman procedure to obtain a path integral
formulation of loop quantum cosmology starting from the Hilbert
space framework.  Quantum geometry effects modify the weight
associated with each path so that the effective measure on the space
of paths is different from that used in the Wheeler-DeWitt theory.
These differences introduce some conceptual subtleties in arriving
at the WKB approximation. But the approximation is well defined and
provides intuition for the differences between loop quantum
cosmology and the Wheeler-DeWitt theory from a path integral
perspective.

\end{abstract}
\pacs{04.60.Kz,04.60Pp,98.80Qc,03.65.Sq}
\maketitle

\section{Introduction}
\label{s1}

In loop quantum cosmology (LQC) \cite{lqcrev,mblrr} quantum effects
become dominant in the Planck era near the big bang and big-crunch
singularities. Because of the underlying quantum geometry of loop
quantum gravity (LQG), the theory inherits a repulsive force. This
force is completely negligible when the curvature is less than, say,
1\% of the Planck scale but then grows dramatically, overwhelming
the classical gravitational attraction and causing a quantum bounce
that resolves the classical singularity \cite{aps1,aps2,aps3}. The
effect is generic in the sense that it holds, for example, in
absence \emph{or} presence of a cosmological constant, anisotropies,
and spatial curvature \cite{bp,ap,awe2,awe3,ewe,apsv}. From a path
integral viewpoint (see, e.g., \cite{swh}), on the other hand, this
stark departure from classical solutions seems rather surprising at
first. For, in the path integral formulation quantum effects usually
become important when the action is comparable to the Planck's
constant $\hbar$ while the action along classical trajectories that
originate or end in the singularity is generically very large. Thus
there is conceptual tension. It is therefore desirable to understand
LQC results from a path integral perspective. The goal of this paper
is to fill this gap at the level of rigor that is common in the
discussion of path integrals.

LQC is formulated in the Hamiltonian framework where one works with
a Hilbert space and operators, appropriately taking into account the
subtleties that arise because of the presence of a Hamiltonian
constraint and the absence of an external time parameter. In the
traditional path integral approach, on the other hand, one generally
\emph{defines} the quantum theory by integrating over the space of
classical geometries, each path being weighted by $e^{iS}$ where $S$
is the Einstein Hilbert action, supplemented by suitable boundary
terms \cite{swh}. These procedures then leads to the conceptual
tension discussed above. To resolve it, we will return to the
original derivation of path integrals \cite{rpf}, where Feynman
began with the expressions of transition amplitudes in the
Hamiltonian theory and \emph{reformulated} them as an integral over
all kinematically allowed paths. But now there is a twist: We are
dealing with a constrained system without external time. In this
case the analog of the transition amplitude is an \emph{extraction
amplitude} ---a Green's function which extracts physical quantum
states from kinematical ones and also provides the physical inner
product between them (see, e.g., \cite{ach1,ach2,hrvw}). If the
theory can be deparameterized, it inherits a relational time
variable and then the extraction amplitude can be re-interpreted as
a transition amplitude with respect to that time \cite{ach2}.
Irrespective of whether this is possible, the basic object that
encodes quantum dynamics is the extraction amplitude and we are led
to start with its expression in LQC and cast it as a path integral.
The final result will be quite similar to the traditional path
integrals based on geometrodynamics \cite{swh}. However, there are
also key differences. As one would expect, these emerge from the
fact that since LQG is based on a quantum theory of geometry, now
one cannot simply start with paths represented by smooth, classical
metrics. It is precisely these differences that lead to large
deviations from the classical behavior in the Planck regime near the
big bang. Thus, it is possible to understand the singularity
resolution of LQC also in the path integral language provided the
choice of paths and the weight assigned to them ---i.e., the
effective measure--- appropriately incorporate effects of quantum
geometry. Our rewriting of the canonical theory in the path integral
framework and the associated WKB approximation is also of interest
in its own right as it is likely to provide new insights and tools
in LQC.

In simplest models one can construct two different path integrals
from the Hilbert space theory. On the one hand, one can choose to
express the transition amplitude as an integral over paths in
\emph{configuration space}. On the other hand, one can consider an
integral over \emph{phase space} paths. In LQC the geometric
operator corresponding to the configuration variable has a discrete
spectrum. As a consequence, if one descends from the Hamiltonian
theory, the construction of a configuration space path integral
leads to a `sum over discrete histories', as in
\cite{ach1,ach2,chn}. This sum resembles the vertex expansion of
spinfoams \cite{newlook} rather than a standard path integral which
features continuous paths with weights given by the exponential of a
phase ($i$ times an action). In this paper, by contrast, we will use
phase space paths and arrive at a standard form of the path
integral. One can then apply usual techniques as for ordinary path
integrals. In particular, we will be able to carry out the saddle
point evaluation of the path integral and use WKB methods.

The paper is organized as follows. We begin in Sec.\ref{s2} by
giving a brief summary of the LQC model we wish to use. In
Sec.\ref{s3} we derive the phase space path integral for this
model. Its saddle point approximation is then studied in
Sec.\ref{s4}. We conclude with a summary and discussion in
Sec.\ref{s5}. In familiar quantum mechanics the WKB
approximation provides the leading term in an $\hbar$ expansion
of the transition amplitude. In Appendix \ref{app1} we recast
that $\hbar$ expansion to extraction amplitudes of constrained
systems. This expansion provides the point of departure for the
WKB approximation in Sec.\ref{s4}. Finally, Appendix\ref{app2}
recalls some technical results from\cite{ach2} that are used in
Sec.\ref{s4}.

\section{Solvable LQC}
\label{s2}

We will focus on the LQC model that has been analyzed in the most
detail \cite{aps1,aps2,aps3,acs}: the k=0, $\Lambda$=0 Friedmann
model with a massless scalar field as a source, which has the advantage
 of being exactly solvable \cite{acs}. However, it would
not be difficult to extend this analysis to allow for a non-zero
cosmological constant \cite{bp,ap}, anisotropies \cite{awe2,awe3}
or to the spatially compact k=1 case \cite{apsv}.

In the FRW models, one begins by fixing a (spatial) manifold $S$,
topologically $\R^3$, Cartesian coordinates $x^i$ thereon, and a
fiducial metric $q^o_{ab}$ given by $q^o_{ab} \dd x^a \dd x^b = \dd
x_1^2 + \dd x_2^2 + \dd x_3^2$. The physical 3-metric $q_{ab}$ is
then determined by a scale factor $a$; \, $q_{ab} = a^2 q^o_{ab}$.
For the Hamiltonian analysis one fixes a cubical fiducial cell $\V$
whose volume with respect to $q^o_{ab}$ is $V_o$ so that its
physical volume is $V = a^3V_o$. The classical gravitational phase
space will be described by the canonical pair
\be \v = \frac{a^3 V_o}{2 \pi G }\,,\qquad  b = -\frac{4 \pi G}{3
V_o}\frac{p_a}{a^2}\, , \ee
where $p_a$ is the conjugate momentum of $a$; their Poisson bracket
is then given by $\{\v,b \}=-2$. The matter phase space is described
by the canonical pair $\p$ and $\pp$, with $\{\p,\pp\}=1$. Note
\emph{two differences from the previous LQC literature}
\cite{acs,ach1,ach2}: there is no $\hbar\gamma$ in the definitions
of $\v$ and $b$ and the momentum conjugate to $\p$ is denoted here
by $p$ rather than $p_{(\p)}$. The first change removes unnecessary
complications in the $\hbar$ expansions one often makes while
working with path integrals while the second just simplifies
notation.

In the quantum theory, the kinematical Hilbert space is a tensor
product $\Hk=\Hkg \otimes \Hkm$ of the gravitational and matter
Hilbert spaces. Elements $\Psi(\v)$ of $\Hkg$ are functions of $\v$
with support on a countable number of points and with finite norm
$||\Psi||^2 := \sum_{\v}\,|\Psi(\v)|^2$. The matter Hilbert space is
the standard one: $\Hkm = L^2(\R, \dd\p)$.

Thus, the kinematic quantum states of the model are functions
$\Psi(\v,\p)$ with finite norm $||\Psi||^2 := \sum_{\v}\, \sint
\dd\p\, |\Psi(\v,\p)|^2$. A (generalized) orthonormal basis in
 $\Hk$ is given by $|\v,\p\ket$ with
\be \label{kin} \bra \v^\prime,\p^\prime\,|\, \v,\p\ket =
\delta_{\v^\prime\v}\,\,\delta(\p^\prime, \p)\, .\ee
A notable feature of kinematics of LQC is that the normalization
involves a Kronecker delta even though $\v$ is a continuous variable
at this stage. Quantum dynamics is encoded in the constraint
equation,
\be \label{qc} -\,\wh{C}\Psi(\v,\p) \equiv \left(-\wh{\pp}^2
+ \Theta \right) \Psi(\v,\p) = 0, \ee
where $\wh{\pp}= - i \hbar \partial_\p $ and $\Theta$ is the
positive, self-adjoint operator acting on $\Hkg$ \cite{kl} given by
\be \label{theta0} \Theta := \frac{3 \pi G}{\lo^2}
\left(\sqrt{|\wh{\v}|}\,\, \wh{\sin \lo b}\,\, \sqrt{|\wh{\v}|}
\right)^2. \ee
Here $\lo$ is related to the `area gap' $\Delta =4\sqrt{3}\pi
\gamma\, \lp^2$ via $\lo^2 = \gamma^2\,\Delta$, where $\gamma$
is the Barbero-Immirzi parameter of LQG, $\wh{\v}$ acts by
multiplication, and $\wh{\e^{i \lo b}}$ by translation:
$\big(\wh{\e^{i \lo b}} \Psi \big)(\v)=\Psi(\v+2 \lo \hbar)$.\,
(Again, there is a small departure from the notation used in
the previous literature \cite{acs,ach1,ach2} in that $\lo^2$
was set to $\Delta$ there. This change will facilitate the
semi-classical considerations.) The explicit form of $\Theta$
is the following second order difference operator
\ba \label{theta} \big(\Theta \Psi\big)(\v) =
- \f{3\pi
G}{4\lo^2} \Big[& \sqrt{|\v(\v+4\lo \hbar)|}\, |\v+2\lo \hbar|\,
\Psi(\v+4\lo \hbar)\,-\, 2\v^2\Psi(\v)\nonumber\\
&+\, \sqrt{|\v(\v-4\lo \hbar)|}\, |\v-2\lo \hbar|\, \Psi(\v-4\lo \hbar)\, \Big]\, .
\ea
From (\ref{theta}) one can see that the space of solutions to the
quantum constraint is naturally decomposed into sectors in which the
wave functions have support on specific `$\v$-lattices' \cite{aps2}.
Furthermore these sectors are preserved by a complete set of
physical observables that is of direct physical interest. Thus there
is superselection and in each superselected sectors the
configuration variable $\v$ assumes discrete values. For
definiteness, we will restrict ourselves to the lattice $\v = 4n\lo
\hbar$ where $n$ is an integer. On this sector $\v$ resembles the
momentum variable of a particle on a circle, whence the conjugate
variable $b$ now lies in a bounded interval $(0, \pi/\ell_o)$.

Solutions to the constraint equation, as well as their inner
product, can be obtained through the group averaging procedure.
Given a state $|\Psi_{\rm kin} \ket$ in the kinematical space $\Hk$,
a physical state $|\Psi_{\rm phys} )$ (i.e. a solution to the
constraint equation) is given by:
\be \label{phy0} |\Psi_{\rm phys} ) = \sint \dd{\alpha} \,
\e^{\frac{i}{\hbar} \alpha \wh{C}} \, |\Psi_{\rm kin} \ket \ee
(see, e.g., \cite{ach2}). Here, we have introduced the $\hbar$
factor in (\ref{phy0}) for later convenience. (Since the dimensions
of the constraint are $[\wh{C}]= \text{M} \text{L}^3$, $\alpha$ has
dimensions of
$\text{L}^{-2}$.) A Green's function%
\footnote{We will actually restrict ourselves to the `positive
frequency part' as in \cite{ach1,ach2}, so that there is an implicit
$\theta(\hat{\pp})$ factor in (\ref{phy1}) where $\theta$ is the
unit step function. We do not write it explicitly just to avoid
unnecessary proliferation of symbols.}
for the above transformation is then given by \cite{ach1,ach2}
\be \label{phy1} \G(\v_f,\p_f; \v_i,\p_i):= \sint \dd{\alpha} \,\,
\bra \v_f,\p_f | \, \e^{\frac{i}{\hbar} \alpha \wh{C}}  \,| \v_i,
\p_i \ket\  , \ee 
in terms of which (\ref{phy0}) can be written as
\be \Psi_{\rm phys}(\v,\p) = \sum_{\v'}\, \sint \dd{\p'}\, \G(\v,\p;
\v',\p') \,\Psi_{\rm kin}(\v',\p'). \ee
In other words, $\G$ gives the matrix elements of the `extractor'
that extracts a physical state from every (suitably regular)
kinematical one. Therefore, it will be referred to as the
\emph{extraction amplitude}.

The inner product between two physical states $|\Phi_{\rm phys} )$
and $|\Psi_{\rm phys} )$ is defined as follows. Let $|\Phi_{\rm kin}
\ket$ and $|\Psi_{\rm kin} \ket$ be kinematical states such that
under the extraction map defined by Eq. (\ref{phy0}) they get mapped
to the given physical states. The physical inner product is then
defined by the action of the `bra' $( \Phi_{\rm phys} |$ on the
`ket' $|\Psi_{\rm kin} \ket$, or equivalently,
\ba (\Phi_{\rm phys},\, \Psi_{\rm phys}) & := & \bra \Phi_{\rm kin}|
\sint \dd{\alpha} \, \e^{\frac{i}{\hbar} \alpha \wh{C}} \,
|\Psi_{\rm kin} \ket \nonumber \\
& = & \sum_{\v,\, \v'}\, \sint \dd{\p}\, \dd{\p'}\,
\bar{\Phi}_{\rm kin}(\v,\p) \G(\v,\p; \v',\p') \Psi_{\rm kin}(\v',\p').
\ea
Thus, in the `timeless' framework without any deparametrization, all
the information of the quantum dynamics is encoded in the extraction
amplitude $\G(\v_f,\p_f; \v_i,\p_i)$. We will start by finding a
path integral expression for this function.

\emph{Remark:} Detailed analysis shows that $\phi$ is a viable
relational time variable \cite{aps1,aps2,aps3} and one can use
it to deparameterize the theory. When this is done, the
extraction amplitude $\G(\v_f,\p_f; \v_i,\p_i)$ can also be
regarded as the amplitude for a transition from $\v_i$ at
`time' $\p_i$ to $\v_f$ and `time' $\p_f$ \cite{ach2}.  Even
though the deparameterized theory is closer to more familiar
path integrals where one also computes transition amplitudes,
in the main body of this paper we will work in the timeless
framework. Although the two are equivalent, the timeless
framework is technically simpler because it leads to a path
integral that directly involves matrix elements of $\Theta$; in
the deparameterized framework it would involve matrix elements
of $\sqrt\Theta$ which are much more complicated \cite{ach2}.

\section{Phase space Path Integral}
\label{s3}

In ordinary quantum mechanics, path integrals provide an expression
for the matrix elements of the evolution operator. Feynman
\cite{rpf} first derived it from the canonical theory by writing the
evolution as a composition of $N$ infinitesimal ones and inserting
complete basis between these infinitesimal evolution operators. He
then arrived at a `discrete time' path integral expression; the
continuum path integral was found by taking the limit $N \to
\infty$.

In the timeless framework there is no `evolution' operator ---all we
have is the constraint equation and its solutions--- and the
extraction amplitude $\G(\v_f,\p_f; \v_i,\p_i)$ replaces the
transition amplitude. Thus, our task is to construct a path integral
expression of this object. The idea is to mimic the standard Feynman
construction \cite{rpf} but now using the operator
$\e^{\frac{i}{\hbar} \alpha \wh{C}}$ that appears on the integrand
of Eq. (\ref{phy1}) for the `evolution' operator, and then
performing the $\alpha$ integration.

Let us be more specific. The integrand of (\ref{phy1}) can be
thought of as a matrix element of the fictitious evolution operator
$\e^{\frac{i}{\hbar} \alpha \wh{C}} $. One can regard  $\alpha
\wh{C}$ as playing the role of a (purely mathematical) Hamiltonian,
the evolution time bring unit, i.e. $\e^{\frac{i}{\hbar} \alpha
\wh{C}} = \e^{\frac{i}{\hbar} t \wh{H}} $ with $\wh{H}=\alpha
\wh{C}$ and $t=1$. We now decompose this fictitious evolution into
$N$ evolutions of length $\epsilon = 1/N$: $\e^{\frac{i}{\hbar} t
\wh{H}} = \prod_{n=1}^N \e^{\frac{i}{\hbar} \epsilon \wh{H}}$. By
inserting complete basis of the form ${\bf 1} = \sum_{v} \sint
\dd{\p} |\v,\p \ket \bra \v,\p |$ in between each factor we get
\ba \label{exact}
 \bra
\v_f,\p_f | \,e^{\frac{i}{\hbar} \alpha \wh{C}} \, | \v_i, \p_i \ket
= \sum_{\v_{N-1},...,\v_1} \sint \dd \p_{N-1} \ldots \dd \p_1 &
 \bra \v_{N},\p_{N} |\, e^{\frac{i}{\hbar} \epsilon \alpha
 \wh{C}} | \, \v_{N-1},\phi_{N-1}  \ket \ldots \nonumber \\
& \ldots \bra \v_{1},\p_{1} |\, e^{\frac{i}{\hbar} \epsilon
\alpha \wh{C}} | \, \v_{0},\phi_{0}  \ket,
\ea
where $\bra \v_{N},\p_{N} | \equiv \bra \v_{f},\p_{f} |$ and $| \,
\v_0,\phi_0 \ket \equiv | \, \v_i,\phi_i \ket $.

Let us concentrate on the $n$-th term appearing in (\ref{exact}).
Notice that since the constraint is a sum of two commuting pieces
that act separately on $\Hkm$ and $\Hkg$, one has the following
factorization:
\be \label{afactor} \bra \v_{n+1},\p_{n+1} |\, \e^{\frac{i}{\hbar}
\epsilon \alpha \wh{C}} \,|  \v_n,\phi_n \ket  =  \bra \p_{n+1} |\,
\e^{\frac{i}{\hbar} \epsilon \alpha \wh{\pp}^2}\, |  \phi_n \ket
\bra \v_{n+1} |\, \e^{-\frac{i}{\hbar} \epsilon \alpha \Theta}\, |
\v_n \ket. \ee
The scalar field factor can easily be evaluated by inserting a
complete basis in $\pp$,
\be \label{aphi} \bra \p_{n+1} |  \e^{\frac{i}{\hbar} \epsilon
\alpha \wh{\pp}^2 }\,   | \p_n \ket = \sint \frac{\dd{\pp_n}}{2 \pi}
e^{\frac{i}{\hbar} \pp_n(\phi_{n+1}-\phi_n)+\frac{i}{\hbar} \epsilon
\alpha  \pp_n^2}. \ee
The gravitational factor in (\ref{afactor}) is less trivial to
compute. As usual in the path integral construction, we will
take $N \gg 1$  ($\epsilon \equiv 1/N \ll 1$) and use an
expansion in $\epsilon$ to compute this term:
\be  \label{agrav1}
 \bra \v_{n+1} | e^{-\frac{i}{\hbar} \epsilon \alpha \Theta} |  \v_n \ket
=  \delta_{\v_{n+1},\v_n}-\tfrac{i}{\hbar} \epsilon \alpha \bra
\v_{n+1} | \Theta |  \v_n \ket + \mathcal{O}(\epsilon^2),\ee
where the matrix element of $\Theta$ can be obtained from Eq.
(\ref{theta}) and is given by
\be \label{metheta} \bra \v_{n+1} | \Theta |  \v_n \ket = -
\frac{3\pi G}{4 \lo^2}\, \sqrt{|\v_n \v_{n+1}|}\,\frac{(\v_n +
\v_{n+1})}{2} (\delta_{\v_{n+1},\v_n+4\lo} - 2\delta_{\v_{n+1},\v_n}
+ \delta_{\v_{n+1},\v_n-4\lo}). \ee 
(There are several equivalent ways of writing this matrix element.
Here we chose one that is symmetric in $\v_n$ and $\v_{n+1}$.) As in
usual path integral constructions, we now bring-in $b$, the
conjugate variable to $\v$. This can be done  through the identity
\be \delta_{\v',\v}=\frac{\lo}{\pi} \sint^{\pi/\lo}_{0} \dd b\,\,
\e^{-\frac{i}{2\hbar} b(\v'-\v)}, \ee
which, when used to rewrite the Kronecker deltas appearing in Eqs.
(\ref{agrav1}) and (\ref{metheta}), leads to the following
expression for (\ref{agrav1}):
\ba \bra \v_{n+1} &|&\!e^{-\frac{i}{\hbar} \epsilon \alpha
\wh{\Theta}}\,\,\, | \v_n \ket \nonumber\\
&=&\frac{\lo}{\pi} \sint_0^{\pi/\lo} \dd b_{n+1} e^{-\frac{i}{\hbar}
b_{n+1}(\v_{n+1} - \v_{n})/2} \left[1 - \frac{i}{\hbar}\epsilon
\alpha \frac{3 \pi G}{\lo^2} \sqrt{\v_{n} \v_{n+1}} \frac{\v_{n} +
\v_{n+1}}{2} \sin^2(\lo b_{n+1})\right]\, +\,
\mathcal{O}(\epsilon^2)
\nonumber \\
& = & \frac{\lo}{\pi} \sint_0^{\pi/\lo} \dd b_{n+1}
e^{-\frac{i}{\hbar} b_{n+1} (\v_{n+1} - \v_{n})/2 - i\epsilon \alpha
\frac{3 \pi G}{\lo^2} \sqrt{\v_{n} \v_{n+1}} \frac{\v_{n} +
\v_{n+1}}{2} \sin^2(\lo b_{n+1})} \, + \, \mathcal{O}(\epsilon^2)
\label{agrav2}. \ea
By combining (\ref{afactor}), (\ref{aphi}) and (\ref{agrav2}), the
amplitude  (\ref{exact}) takes the form
\ba \label{pint1} \bra \v_f,\p_f | \,e^{i \alpha \wh{C}} \, | \v_i,
\p_i \ket  = \sum_{\v_{N-1},...,\v_1} &(\f{\lo}{\pi})^N \sint \dd
b_{N} \ldots \dd b_1 \sint \dd \p_{N-1} \ldots \dd \p_1\,\,
\times\,\, \nonumber\\
&(\f{1}{2\pi})^N\, \sint \dd \pp_{N} \ldots \dd \pp_1
\,\e^{\frac{i}{\hbar} S_N} + \mathcal{O}(\epsilon^2), \ea
where
\ba \label{DiscS} S_N=  \epsilon \sum_{n=0}^{N-1}\, \Big[ \pp_{n+1}
\frac{\phi_{n+1}-\phi_{n}}{\epsilon} &-& \frac{b_{n+1}}{2}
\frac{\v_{n+1}-\v_{n}}{\epsilon} \nonumber \\
&+& \alpha \big(\pp_n^2-\frac{3 \pi G}{\lo^2} \sqrt{\v_{n} \v_{n+1}}
\frac{\v_{n} + \v_{n+1}}{2} \sin^2(\lo b_{n+1})\big) \Big] \ea
can be heuristically regarded as a `discrete-time action'. Thus,
following Feynman, we have obtained an approximate expression of the
extraction amplitude as a sum over phase space paths, approximation
becoming better and better as $\epsilon$ shrinks and $N$ grows.

Already, before taking the $N\rightarrow \infty$ limit,  we can see
two important differences from the more familiar path integrals.
First, whereas in the classical theory the variables $\v$ and $b$
can take all real values, our paths in $\v$ take only discrete
values and the paths in $b$ are bounded; the allowed values of $\v$
and $b$ are dictated by spectrum of the corresponding operators on
the superselected sector we began with. Second, we find that the
action is not just a discretization of the classical action, but
includes quantum gravity corrections (namely, the so-called
`holonomy type corrections' encoded here in the $\sin$ term). What
is the origin of these differences? After all, we just mimicked the
Feynman construction \cite{rpf} to arrive at (\ref{DiscS}) starting
from the Hamiltonian theory. Recall however, that Feynman began with
the standard Schro\"odinger representation of quantum mechanics. In
LQG, because of the underlying diffeomorphism invariance the
analogous representation is not viable \cite{lost}. This difference
descends to LQC where kinematics is based on a `polymer'
representation that is unitarily \emph{inequivalent} to the
Schr\"odinger representation \cite{abl}. In the polymer
representation, the quantum constraint operator of the
Wheeler-DeWitt theory ---which would have been analogous to the
Hamiltonian operator Feynman used--- fails to be well-defined.
Defining a viable constraint operator on the kinematical Hilbert
space of LQC requires an appropriate incorporation of quantum
geometry underlying LQG \cite{aps3}. And this is directly
responsible for the two key differences mentioned above.

The final step in the path integral construction involves taking the
limit $N \to \infty$. In this limit one typically performs the
substitutions  $\epsilon \sum_{n=0}^{N} \to \sint \dd \tau$,\,\,\,
$(\phi_{n+1}-\phi_{n})/\epsilon \to d \phi / d \tau$, etc,  to
obtain a continuum (in time) action.  However, due to the discrete
nature of $\v$, it is not possible to interpret the
$(\v_{n+1}-\v_{n})/\epsilon$ as a derivative. Therefore, we are led
to carry out an `integrating by parts,' rewriting this term as
\be \epsilon \sum_{n=0}^{N-1} \left[ - \frac{b_{n+1}}{2}
\frac{\v_{n+1}-\v_{n}}{\epsilon} \right] =  \epsilon
\sum_{n=0}^{N-1} \left[ \frac{\v_{n}}{2}
\frac{b_{n+1}-b_{n}}{\epsilon} \right] + \frac{1}{2} (b_1 \v_0 - b_N
\v_N).
 \ee
Now the formal limit can be carried out at the same level of
precision as is common in path integrals and one obtains
\be \label{pi1} \G(\v_f,\p_f; \v_i,\p_i)  = \sint \dd \alpha\,\,
\sint [\mathcal{D}\v_q(\tau)]\,\, [\mathcal{D}b_q(\tau)]\,\,
[\mathcal{D}\pp(\tau)]\,\, [\mathcal{D}\p(\tau)]
\,\,\,\e^{\frac{i}{\hbar} \bar{S}}, \ee
where
\be \label{qia1} \bar{S}=\int_{0}^{1}d\tau\left(\pp
\dot{\phi}+\frac{1}{2} \v \dot{b}-\alpha \left(\pp^2-3\pi G \v^2
\frac{\sin^2\lo b}{\lo^2}\right) \right) - \frac{1}{2} (\v_f b_f -
\v_i b_i) \ee
and the subscript $q$ emphasizes the fact that we have taken the
formal limit of a sum that included only `quantum paths' in the
geometrical sector. Thus, the integral is over paths $\v(\tau)$
taking only the values $\v = 4n\lo \hbar$ and the paths $b(\tau)$
take values in the range $(0, \pi/\lo)$. Notice that the classically
singular paths where $b$ is divergent are \emph{excluded}, which
reflects the fact that the quantum dynamics is free of
singularities. However, precisely because of the unusual restriction
on the domain of integration, this path integral is not of the usual
type. The path integral (\ref{pi1}) actually resembles  the `sum
over histories' of \cite{ach2}, where one sums over the same family
of $\v$-paths. Indeed, by performing the $b_q$ integral in
(\ref{pi1}) one can recover the sum over histories expansion of
\cite{ach2}.

But it turns out that we can express this path integral also in a
more familiar form using a clever trick from path integral framework
for a particle in a circle \cite{kleinert}. The trick is to use the
identity
\be\sum_{m \in \mathbb{Z}}\int_{0}^{2\pi}\dd
\theta\,f(\theta,m)\,e^{i m \theta}=\int_{-\infty}^{\infty}\dd x
\int_{-\infty}^{\infty}\dd\theta\, f(\theta,x)\, e^{i x \theta}\ee
which holds for any continuous $f(\theta,x)$ with a $2 \pi$ period
in $\theta$ to convert the discrete sum over $m$ to a continuous
integral over $x$. Let us then return to the finite $N$ or
`discrete-time' path integral (\ref{pint1}) and use this identity.
Then, we are led to rewrite each sum over $\v_n$ and integral over
$b_n$ appearing in (\ref{pint1}) as
\be \f{\pi}{ \lo} \sum_{\v_n} \int_{0}^{\pi/\lo} \dd b_n \ldots
\quad\to\quad \int_{-\infty}^{\infty}\dd \v_n
\int_{-\infty}^{\infty} \dd b_n \ldots\ee
(this is done for $n=1, \ldots, N-1$; the integral over $b_N$
remains unchanged). The allowed paths now take value over the whole
classical phase space: $\v$ is no longer restricted to be discrete
nor is $b$ required to be bounded. As in standard path integral
discussions, it is then possible to take the formal limit
$N \to \infty$. We obtain%
\footnote{ Note that the integral over $\alpha$ is an ordinary one
variable integral. It can nevertheless we reinterpreted as an
integral over \emph{all} possible values $\alpha(\tau)$ together
with the a gauge fixing condition  $\dd \alpha /\dd\tau = 0$. Doing
so allows one to rewrite the path integral with any other gauge
fixing condition; see for instance \cite{marolfpi}.}
\be \label{pi} \G(\v_f,\p_f; \v_i,\p_i)  = \sint \dd \alpha\, \sint
[\mathcal{D}\v(\tau)]\,\, [\mathcal{D}b(\tau)]\,\,
[\mathcal{D}\pp(\tau)]\,
[\mathcal{D}\p(\tau)]\,\,\,\e^{\frac{i}{\hbar} S}, \ee
where
\be \label{qia}
S=\int_{0}^{1}d\tau\left(\pp \dot{\phi}-\frac{1}{2}b
\dot{\v}-\alpha \left(\pp^2-3\pi G \v^2 \frac{\sin^2\lo
b}{\lo^2}\right) \right). \ee
While this is the same action as before, \emph{one now integrates
over all trajectories in the classical phase space as in usual path
integrals.} In particular, the trajectories that make the action
stationary lie in the domain of integration (which, by contrast,
were not included in the `quantum paths' of the previous path
integral). We can therefore use all the standard techniques such as
the saddle point approximation. Thus, while they are (formally)
equivalent, this second form (\ref{pi}) of the path integral is much
more convenient in practice, particularly to address the issues we
began with in Sec.\ref{s1}. This is the path integral form of the
extraction amplitude we were seeking.

Since we now integrate over all paths in the classical phase space,
in particular, the paths are allowed to go through the classical
singularity. How can then we see the singularity resolution in this
setting? The answer is that the paths are not weighted by the
standard FRW action but by a `polymerized' version of it which still
retains the memory of the quantum geometry underlying the
Hamiltonian theory. As we will see, this action is such that a path
going through the classical singularity has negligible contribution
whereas bouncing trajectories give the dominant contribution.

\emph{Remark:} There are other systems in which the passage
from the Hamiltonian quantum theory to a path integral results
in an action that has $\hbar$-corrections. Perhaps the simplest
example is that of a non-relativistic particle on a curved
Riemannian manifold for which the standard Hamiltonian operator
is simply $\hat{H} = - (\hbar^2/2m) g^{ab} \nabla_a\nabla_b$.
Quantum dynamics generated by this $\hat{H}$ can be recast in
the path integral form following the Feynman procedure
\cite{rpf}. The transition amplitude is then given by
\cite{bdw}
\be \langle q,t| q',t'\rangle = \sint \mathcal{D}[q(\tau)]\,
e^{\f{i}{\hbar} S} \ee
with
\be S = \sint \dd \tau\, (\f{m}{2} g_{ab}\dot{q}^a \dot{q}^b +
\f{\hbar^2}{12m} R)\ee
where $R$ is the scalar curvature of the metric $g_{ab}$.
Extrema of this action are not the geodesics one obtains in the
classical theory but rather particle trajectories in a
$\hbar$-dependent potential. The two sets of trajectories can
be qualitative different.

\section{Saddle point approximation}
\label{s4}

As we saw in Sec.\ref{s2}, the extraction amplitude encodes the
entire content of quantum dynamics. However, in practice, it is
difficult to work with the series (\ref{pint1}) or evaluate its
limit (\ref{pi}). In quantum mechanics and quantum field theory the
steepest descent approximation is a powerful practical tool to
calculate leading contributions to the transition amplitude in an
$\hbar$ expansion. In particular, this approximation provides the
much needed intuition on when quantum corrections are dynamically
important and when they are not. In Appendix \ref{app1} we recast
this $\hbar$ expansion in a form suitable for the extraction
amplitude of constrained systems. We will now use those results to
obtain the leading term using a saddle point approximation.

In this approximation, the extraction amplitude (\ref{pi}) is given
by
\begin{equation}\label{spa}
\G(\v_f,\p_f; \v_i,\p_i) \sim \left(\det{\delta^2 S}|_0
\right)^{-1/2}\,\, \e^{\frac{i}{\hbar} S|_0 }\,.
\end{equation}
Here $S|_0$ is the action evaluated along the trajectory extremizing
the action with initial and final configuration points fixed. For
given initial and final points, there exist in general two
trajectories joining them, one with positive and the other with
negative $\pp$ values. As pointed out in Sec.\ref{s2}, following the
logic spelled out in \cite{ach2} here we restrict ourselves to the
`positive frequency' branch, and so only the $\pp>0$ trajectory gets
picked. We will evaluate the phase factor in Sec.\ref{s4.1}. The
prefactor $\left(\det{\delta^2 S}|_0 \right)^{-1/2}$ represents a
formal infinite dimensional determinant which we will evaluate in
Sec.\ref{s4.2}. In Sec.\ref{s4.3} we compare the resulting
approximate amplitude to the exact one, computed numerically.

Before proceeding with these calculations, we would like to
point out a conceptual subtlety. In ordinary quantum mechanics,
the steepest descent approximation provides the leading term in
the transition amplitude in an $\hbar$ expansion. In our case,
the action $S$ that features in the path integral (\ref{pi})
itself depends on $\hbar$ through $\lo\sim \sqrt{\gamma^3 \hbar
G}$, while the $\hbar$ expansion of Appendix \ref{app1} assumes
that the action does not change as $\hbar$ tends to zero.
Therefore, to directly apply the result of Appendix \ref{app1},
now we have to take the limit $\hbar \to 0$ while keeping $\lo$
fixed. Hence we will obtain the leading term in the extraction
amplitude in the approximation $\hbar \to 0$,\, $\gamma \to
\infty$ keeping $\gamma^3 \hbar$ fixed. To emphasize this
subtlety, \emph{we will use inverted commas, as in `classical
limit' and `semi-classical approximation' while referring to
this limit.} Let us briefly explore the meaning of this limit.
In classical general relativity, $\gamma\to \infty$ corresponds
to ignoring the new term in the Holst action for general
relativity, in comparison with the standard Palatini term. What
about the `semi-classical' approximation? Eigenvalues of the
volume operator are given by $(8\pi G\lo \hbar)n$ where $n$ is
a non-negative integer. Therefore, in the `semi-classical
limit' the spacing between consecutive eigenvalues goes to zero
and $\v$ effectively becomes continuous as one would expect.
Finally, states that are relevant in this limit have large $n$,
just as quantum states of a rigid rotor that are relevant in
the semi-classical limit have large $j$.

\subsection{The Hamilton-Jacobi function $S|_0$}
\label{s4.1}

To calculate the $S|_0$ term, we need to solve the equations of
motion obtained from the action (\ref{qia}), then evaluate the
action along those trajectories, and finally express the result
in terms of the given initial and final points.  The (positive
frequency) trajectories which solve the equations of motion can
be written in terms of two integration constants, $\v_B$ and
$\p_B$, as
\ba
\v(\p) & = & \v_B \cosh(\sqrt{12 \pi G}(\p-\p_B)) ,
\label{solns1} \\
b(\p) & = & \frac{2\; \text{sign}(\v_B)}{\lo} \tan^{-1}(e^{-\sqrt{12 \pi
G}(\p-\p_B)}) \label{solns2}. \ea
These solutions have several interesting features.\\
(i) As seen from the $\cosh$ dependence of the volume, these
trajectories represent bouncing universes, with $\p_B$ and $\v_B$
giving the scalar field and volume values at the bounce point. The
minimum volume $\v_B$ is related with  the scalar field momentum
$\pp$ by $|\v_B|={2 \lo \pp}/{\sqrt{12 \pi G}}$. Note that if $\v_B$
is positive (resp. negative), then $\v(\p)$ remains positive (resp.
negative) for all $\p$. For concreteness we will focus on
trajectories with positive $\v_B$.\\
(ii) $\v(\p)$ can vanish only on the trajectory with $\v_B=0$ i.e.,
$\v(\p) = 0$ for all $\p$. Thus if we begin with the initial state
$\v_i\not=0,\p_i$, there is no (real) `classical' trajectory at
all with $\v_f=0$ for any value of $\p_f$.\\
(iii) Whereas in general relativity \emph{all} trajectories begin at
the big-bang ---they all tend to $\v =0$ as $\p \to -\infty$--- it
is obvious from (\ref{solns1}) that all our trajectories tend to $\v
\to \infty$ in this limit (except for the trajectory
$\v(\p) =0\; \forall \p$).\\
(iv) Recall that in full LQC, states which are sharply peaked at a
low curvature configuration for large values of $\p$ remain sharply
peaked on certain `effective trajectories' for all $\p$ \cite{aps3}.
These are among solutions (\ref{solns2}).\\
(v) The relation between $\v$ and $\p$ given in Eq. (\ref{solns1})
coincides with the expression for the expectation value of the
volume operator at a given scalar field value $\p$ in \emph{any}
quantum state of LQC \cite{acs}.

\begin{figure}[h]
\centering
\includegraphics[width=0.35\textwidth]{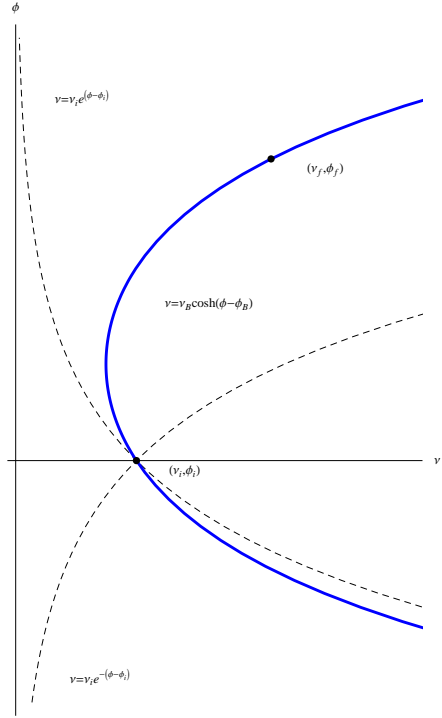}
\caption{For fixed ($\v_i,\p_i)$, the (dashed) curves  $\v_f= \v_i\,
\e^{\pm \sqrt{12 \pi G} (\p_f-\p_i)}$ divide the $(\v_f,\p_f)$
plane into four regions. For a final point in the
upper or lower quarter, there always exists a real trajectory joining
the given initial and final points (as exemplified by the thick line).
If the final point lies on the left or right quarter, there is no real
solution matching the two points. The action becomes imaginary and one
gets an exponentially suppressed amplitude.}
\label{figregions}
\end{figure}

Evaluation of the action along these solutions can be greatly
simplified if one integrates by parts the term $-\int_{0}^{1}d\tau\,
\frac{1}{2}b \dot{\v}$ in (\ref{qia}). Then, using the equations of
motion, the terms $ \frac{1}{2}\dot{b} \v$ and $\pp \dot{\phi}$
cancel each other and the action evaluated along the solutions is
just given by only the boundary term,
\be \label{S01} S|_{0}= \frac{1}{2}\left(\v_i b_i-\v_f b_f \right).
\ee
To express $S|_0$ in terms of initial and final configuration
variables, we need to solve for the constants $\v_B$ and $\p_B$
in terms  of $\v_f,\p_f; \v_i,\p_i$. Without loss of generality
we can take $\p_i=0$ and $\p_f=\varphi$ (by setting $\varphi=
\p_f-\p_i$ at the end, one recovers the general case).  Then we
are led to solve the equations
\ba \v_i &  = & \v_B \,\cosh(-\sqrt{12 \pi G}\p_B) \\ \label{xi}
\v_f & = & \v_B \cosh(\sqrt{12 \pi G}(\varphi-\p_B)) \label{xf}, \ea
for $\v_B$ and $\p_B$ in terms of the initial and final data:
\ba \e^{\sqrt{12 \pi G}\p_B}  & = &  \sqrt{\frac{\e^{\sqrt{12
\pi G}\varphi}
-\v_f/\v_i}{-\e^{-\sqrt{12 \pi G}\varphi}+\v_f/\v_i}}
\label{phi0} \\
\v_B & = & \frac{\v_i}{|\sinh (\sqrt{12 \pi G}\varphi)|} \sqrt{\left(
\e^{\sqrt{12 \pi G}\varphi}- \f{\v_f}{\v_i} \right)
\left(-\e^{-\sqrt{12 \pi G}\varphi}+\f{\v_f}{\v_i} \right)}
\,\label{x0}. \ea
Clearly, $\v_B,\p_B$ are real for any given initial
configuration $(\v_i,\p_i)$ if and only if the final
configuration satisfies
\be \label{region} \e^{-\sqrt{12 \pi G} |\varphi|}\, < \,
\f{\v_f}{\v_i} \,< \,\e^{\sqrt{12 \pi G}|\varphi|}. \ee
This is the necessary and sufficient condition for the
existence of real trajectories. Let us first focus on the
`classically' allowed region (\ref{region}). Using (\ref{solns2})
to express $b_i$ and $b_f$ appearing in (\ref{S01}) in terms of
the initial and final data $(\v_f,\p_f; \v_i,\p_i)$ we obtain
the desired expression of the Hamilton-Jacobi function:
\be \label{S02} S|_0=  \frac{\v_i}{\lo} \tan^{-1}\left(
\sqrt{\frac{\e^{\sqrt{12 \pi G} \varphi}-\v_f/\v_i}{-\e^{-\sqrt{12
\pi G}\varphi}+\v_f/\v_i}}\right)- \frac{\v_f}{\lo} \tan^{-1}\left(
\e^{-\sqrt{12 \pi G}\varphi} \sqrt{\frac{\e^{\sqrt{12 \pi
G}\varphi}-\v_f/\v_i} {-\e^{-\sqrt{12 \pi
G}\varphi}+\v_f/\v_i}}\right), \ee
where $\varphi=\p_f-\p_i$. The `classically' allowed region consists
of the upper and lower quarters in Fig. \ref{figregions}. For
$\v_f,\p_f$ in these two quarters, $S_0$ is real and thus the
amplitude (\ref{spa}) has an oscillatory behavior. Outside these
regions the action becomes imaginary and one gets an exponentially
suppressed amplitude. Thus, the situation is analogous to that in
quantum mechanics.

For completeness, let us now discuss the case where the final
point lies in the `classically' forbidden region, this is to
say the situation where the boundary data satisfy
\be \label{region2} \f{\v_f}{\v_i} < \e^{-\sqrt{12 \pi G} |\varphi|
} \quad \text{or} \quad \f{\v_f}{\v_i} > \e^{\sqrt{12 \pi G}
|\varphi| }. \ee
For concreteness let $\v_i$ be positive as in
Fig.{\ref{figregions} but now there is no restriction on the
sign of $\v_f$. To find extrema of the action that join the
initial and final configurations satisfying (\ref{region2}), we
can follow the semi-classical procedure used to calculate
tunneling amplitudes in familiar systems and allow paths with
imaginary momenta. Let us define
\be \tilde{b} = i b , \quad \tilde{p}= i p , \quad \tilde{\alpha}= i
\alpha, \quad \tilde{S}= i S. \ee Eq. (\ref{qia}) then implies
\be \label{stilde}
\tilde{S}=\int d\tau\left(\tilde{p} \dot{\phi}-\frac{1}{2}\tilde{b}
\dot{\v}+\tilde{\alpha} \left(\tilde{p}^2-3\pi G \v^2 \frac{\sinh^2\lo
\tilde{b}}{\lo^2}\right) \right).
\ee

We now consider the case when the tilde quantities are real and
compute the stationary trajectories of $\tilde{S}$. The `positive
frequency' (i.e. $\tilde{p}>0$) trajectories are parameterized by
two integrations constants, $\v_o$ and $\phi_o$, and take the form
\ba
\v(\p) & = & \v_o \sinh(\sqrt{12 \pi G}(\p-\p_o)) ,  \\
\tilde{b}(\p) & = & \frac{2\; \text{sign}(\v_o)}{\lo}
\tanh^{-1}(e^{-\sqrt{12 \pi
G}|\p-\p_o|}). \label{btilde}
\ea
They represent universes that go through a singularity at $\p=\p_o$,
where the volume vanishes and $\tilde{b}$ diverges. As in the
`classically' allowed region we have $|\v_o|={2 \lo \tilde{p}}/
{\sqrt{12 \pi G}}$. In terms of the initial and final data, the
integration constants are
\ba
\e^{\sqrt{12 \pi G}\p_o} & = &  \sqrt{\frac{\e^{\sqrt{12 \pi G}\varphi}
-\v_f/\v_i}{\e^{-\sqrt{12 \pi G}\varphi}-\v_f/\v_i}}  \\
\v_o & = &  \frac{|\v_i| \,\text{sign}(\v_f-\v_i)} {\sinh
(\sqrt{12 \pi G}\varphi)} \sqrt{\left( \e^{\sqrt{12 \pi
G}\varphi} -\f{\v_f}{\v_i} \right) \left(\e^{-\sqrt{12 \pi
G}\varphi}-\f{\v_f}{\v_i} \right)} \,, \ea
which, as expected, take real values in the `forbidden' region
(\ref{region2}). The action can then be evaluated as before.
Although now the paths encounter a divergence in $\tilde{b}$, the
integral (\ref{stilde}) is convergent and given by the tilde version
of (\ref{S01}). (Moreover, the product $\v(\phi) \tilde{b}(\phi)$ is
always finite and vanishes at $\phi = \phi_o$.) For the case when
$\varphi>0$ and $\v_f/\v_i < \e^{-\sqrt{12 \pi G} \varphi}$  the
result is
\be \label{stilde0}  \tilde{S}|_0=  \frac{\v_i}{\lo}
\tanh^{-1}\left( \sqrt{\frac{\e^{-\sqrt{12 \pi G}
\varphi}-\v_f/\v_i}{\e^{\sqrt{12 \pi G} \varphi
}-\v_f/\v_i}}\right) - \frac{\v_f}{\lo} \tanh^{-1}\left(
\e^{\sqrt{12 \pi G} \varphi} \sqrt{\frac{\e^{-\sqrt{12 \pi G}
\varphi}-\v_f/\v_i} {\e^{\sqrt{12 \pi
G}\varphi}-\v_f/\v_i}}\right). \ee
Similar expressions hold for other regions. For instance, in the
$\varphi>0$,\, $\v_f/\v_i > \e^{\sqrt{12 \pi G} \varphi }$ case the
action takes the same form, except that the arguments of the
$\tanh^{-1}$ functions are the reciprocals of the ones appearing in
(\ref{stilde0}).

In all cases, $\tilde{S}|_0$ is negative; the extraction
amplitude is exponentially suppressed for paths in the
`classically forbidden' regions. But as we approach the dashed
curves marking the boundary of the `classically' allowed and
forbidden regions, the action $\tilde{S}$ goes to zero. In
particular then, from Fig.{\ref{figregions}} it may appear that
for any given $\v_i$ there is a significant probability of
reaching the singularity $\v_f=0$ for large $\varphi =
\p_f-\p_i$. However, as is common in more familiar systems, the
steepest descent approximation also becomes poor in a
neighborhood of the dashed curves! Indeed, we know from full
LQC that (in the deparameterized framework) the expectation
value of $\widehat{|\v|}$ tends to \emph{infinity} for large
$\varphi$. More generally, plots of the exact extraction
amplitudes in Sec.{\ref{s4.3} will show that the amplitude is
always suppressed in the classically forbidden regions. Thus,
while the steepest descent approximation provides much physical
insight, it is by no means a substitute for the full quantum
theory.

\subsection{$\det{\delta^2 S}|_0 $ and the WKB approximation}
\label{s4.2}

To compute the amplitude $\left(\det{\delta^2 S}|_0
\right)^{-1/2}$, one would need to perform some regularization
in order to deal with the infinite dimensional determinant.
This is can be done in ordinary quantum mechanics or field
theory, and should as well be doable here. We will however take
a different route and calculate this factor by means of the WKB
approximation \cite{guillemin,bates}.

Note that the extraction amplitude $\G(\v_f,\p_f; \v_i,\p_i)$ can be
thought of as a physical state if one takes the initial data as
fixed parameters and the final data as arguments of the
wavefunction: the family of states
\be \Psi_{\v_i,\p_i}(\v_f,\p_f):= \G(\v_f,\p_f; \v_i,\p_i), \ee
parameterized by $\v_i$ and $\p_i$ satisfy the constraint equation
\be \label{ceqwkb} \wh{C}\, \Psi_{\v_i,\p_i} = 0. \ee
The $\hbar$ expansions underlying the desired WKB approximation
are discussed in Appendix \ref{app1}. We begin with the ansatz
for the physical state:
\be \Psi_{\v_i,\p_i}(\v_f,\p_f) = a(\v_f,\p_f; \v_i,\p_i) \,
\e^{\f{i}{\hbar}\, W(\v_f,\p_f; \v_i,\p_i)\,+\, \mathcal{O}(\hbar)}.
\ee
Following the procedure of Appendix \ref{app1}, the imposition
of the constraint equation (\ref{ceqwkb}) to zeroth and first
order in $\hbar$ leads to the following equations for $a$ and
$W$:
\be C(\v_f,\p_f,\partial_{\v_f} W,\partial_{\p_f} W)=0, \quad {\rm
and}\quad \mathcal{L}_X\, a=0 \ee
where
$$C(\v_f,\p_f,b_f,\pp_f)=\pp_f^2-3 \pi G \frac{\sin^2 \lo b_f}{\lo^2}
\v_f^2 $$
is the `effective constraint', and
$$X=\left.\f{\partial C}{\partial p_f}\right|_{p_f=\partial_{q_f} W}
\,\f{\partial}{\partial q_f}$$
is the vector field on configuration space $q_f=(\v_f,\p_f)$
obtained from the Hamiltonian vector field of the constraint.

The first equation is the Hamilton-Jacobi equation and, as
expected, one can check that $S|_0$ given by (\ref{S02}) solves
it. The amplitude $a$ is determined by the second equation
together with the condition
\be a(\v_i,\p_i;\v_f,\p_f)=a(\v_f,\p_f;\v_i,\p_f) \ee
which follows from the fact that
$\bar{\Psi}_{\v_i,\p_i}(\v_f,\p_f)=\Psi_{\v_f,\p_f}(\v_i,\p_i)$:
\be \label{fluct} a= |\v_i^2\,(\e^{\sqrt{12 \pi
G}\varphi}-\f{\v_f}{\v_i})\,\,(-\e^{-\sqrt{12 \pi G}
\varphi}+\f{\v_f}{\v_i})|^{-1/4}  . \ee
This is the factor we identify with $\left(\det{\delta^2 S}|_0
\right)^{-1/2}$. Note that this quantity diverges at $\v_f= \v_i
\e^{\pm \sqrt{12 \pi G}\varphi}$ (dashed lines in Fig.
\ref{figregions}) where the amplitude goes from oscillatory to
exponential decay behavior. Thus, the WKB approximation can be valid
only away from the dashed lines. This simply mirrors what happens in
the WKB approximation in ordinary quantum mechanics.

To summarize, we have succeeded in finding a saddle point
approximation of the path integral as in equation (\ref{spa}).
The determinant factor was not calculated directly but by
matching with the terms of a WKB expansion. Therefore, we will
call the resulting approximate extraction amplitude
$\G^{\text{WKB}}$:
\be \label{awkb} \G^{\text{WKB}}(\v_f,\p_f; \v_i,\p_i) := a\,
\e^{\frac{i}{\hbar} S|_0}, \ee
where $a$ is given by Eq. (\ref{fluct}), $S|_0$ by Eq.
(\ref{S02}) and as before $\varphi= \p_f-\p_i$. We now proceed
to numerically compare this approximate amplitude with the
exact one.

\subsection{Comparison with exact solution}
\label{s4.3}

One of the advantages of the model under study is its solvability
\cite{acs}. In particular, it is possible to obtain a closed form
expression of the extraction amplitude $\G(\v_f,\p_f; \v_i, \p_i)$
\cite{ach2}. This is displayed in Appendix \ref{app2}. We calculated
the exact solution numerically and compared it with the saddle point
approximation obtained in Sec.\ref{s4.2}. We found that there is a
good agreement away from the dashed lines of Fig. \ref{figregions}
which mark the transition between the `classically' allowed region
to the `classically' forbidden one. Along the dashed line, however,
the WKB amplitude diverges and the approximation fails badly just as
in ordinary quantum mechanics.

\begin{figure}[h]
  \begin{center}
      \includegraphics[width=3in]{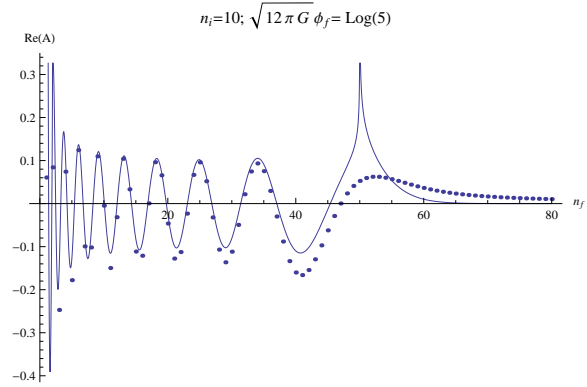}
    \caption{Real parts of the exact and WKB amplitudes are plotted as a
    function of the final volume $n_f := {\v_f}/{4 \lo \hbar}$.
    The exact amplitude (dots) has support on the `lattice'
    $(\v_f-\v_i)/{4 \lo \hbar} \in \Z$. At $n_f=n_i \e^{\sqrt{12 \pi G}
    (\p_f-\p_i)}=50$ there is the transition from oscillatory to exponential
    behavior, and the WKB amplitude (solid line) diverges. (It also diverges
    at $n_f=n_i\, \e^{-\sqrt{12 \pi G} (\p_f-\p_i)}=2$.) Here, $\p_i,\v_i$
    and $\p_f$ are kept fixed: $\p_i=0,\,  n_i := {\v_i}/{4 \lo \hbar}=10$ and
    $\sqrt{12 \pi G}\p_f = \log 5$.}
  \label{p2}
 \end{center}
\end{figure}

\begin{figure}[h]
  \begin{center}
      \includegraphics[width=3in]{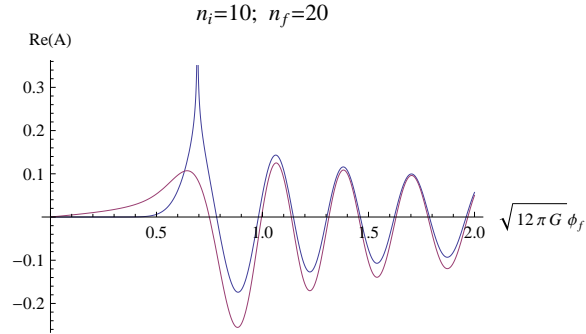}
    \caption{Real parts of the exact and WKB amplitudes are plotted as a
    function of the final scalar field $\p_f$ (dots and solid lines,
    respectively). Here $\p_i,\v_i$ and $\v_f$ are kept fixed: $\p_i=0,
    n_i :={\v_i}/{4 \lo \hbar}=10$ and $n_f :={\v_f}/{4 \lo \hbar}=20$.
    The WKB solution is the curve diverging at $\sqrt{12 \pi G}
    \p_f = \log(\v_f/\v_i)=\log 2$.}
  \label{p3}
 \end{center}
\end{figure}

We illustrate these results in figures \ref{p2}, \ref{p3} and
\ref{p4}. In the first two figures we plot of the real parts of the
exact and WKB amplitudes as a function of $\v_f$ and $\p_f$
respectively, for fixed values of the remaining variables. The exact
amplitude shows a sudden transition from oscillatory to decaying
behavior. If one had access only to the exact result, this behavior
would have seemed rather puzzling. The WKB approximation provides a
physical understanding of this behavior. Thus, not only does the WKB
approximation reproduce the qualitative behavior of the exact
extraction amplitude away from the dashed lines of
Fig.\ref{figregions}, but it anticipates that the dashed lines mark
a boundary between two quite different behaviors of the exact answer
and provides a physical understanding of this difference.

\begin{figure}[h]
  \begin{center}
      \includegraphics[width=3in]{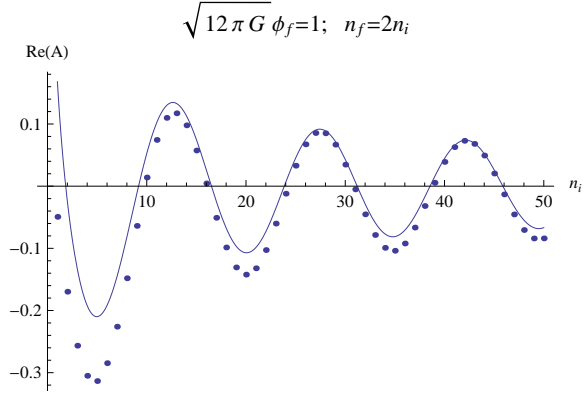}
    \caption{Comparison of the exact and WKB amplitudes for initial and
    final configurations with $\v_f = 2 \v_i$ as a function
    of $n_i := {\v_i}/{4 \lo \hbar}$. As $n_i$ increases,
    the WKB solution (continuous line) becomes closer to the exact
    solution (dots). (The distance between the amplitudes oscillates,
    but overall it decreases.) This calculation was done for
    $\p_i=0$ and $\p_f=1/\sqrt{12 \pi G}$.}
  \label{p4}
 \end{center}
\end{figure}

What can we say regarding the regime of validity of the saddle
point approximation? From the path integral perspective, we
expect it to be valid whenever $S/\hbar \gg 1$. From Eq.
(\ref{S02}), we see that $S|_0$ scales with the volume times a
coefficient which can be interpreted as measuring the departure
from the dashed lines $\v_f=\v_i \e^{\pm \sqrt{12 \pi G}
|\varphi |}$. At these lines $S|_0=0$ and, as we have just
seen, the approximation totally breaks down. As we depart from
these lines, $S|_0$ takes a nonzero value, and its scale is
given by the initial and final volume. For instance, if we keep
the ratio $\v_f/\v_i$ fixed, the action grows linearly with
$\v_i$ and we expect the approximation to improve as $\v_i$
increases. This behavior is indeed observed, an example of
which is display in Figure \ref{p4}. Thus, the standard
expectations on the validity of the WKB approximation are all
borne out.

\section{Discussion}
 \label{s5}

For constrained quantum systems, the extraction amplitude encodes
full quantum dynamics: It enables one to extract physical quantum
states from (suitably regular) kinematical ones \emph{and} define
the physical scalar product between them. In this paper we
considered a solvable model in LQC. Following Feynman, we began with
the expression of the extraction amplitude $\G(\v_f,\p_f; \v_i,
\p_i)$ in the Hilbert space framework and obtained two equivalent
forms of a phase space path integral for it. In the first, one is
led to integrate over paths that are very different from those one
might have expected from the Wheeler-DeWitt theory: The integral is
taken over \emph{quantum paths} defined by the spectrum of the
operators related to the phase space variables. However, by a clever
trick from the quantum theory of a particle on a circle
\cite{kleinert}, this path integral could be reduced to one over
standard classical paths. This form is better suited for
semiclassical considerations and is also closer to that used in the
Wheeler-DeWitt theory. However in \emph{neither} form is the weight
associated with a path given by the Einstein-Hilbert action.
Instead, it is given by an action that includes quantum geometry
corrections. This is the key difference from the Wheeler-DeWitt
theory. Note that such a change of action in the transition from the
Hamiltonian quantum theory to a path integral can occur already in
much simpler systems. For example, for a particle moving on a
Riemannian manifold, dynamics generated by the \emph{standard}
Hamiltonian operator, $\hat{H} = - (\hbar^2/2m)\, g^{ab} \nabla_a
\nabla_b$,  is correctly captured in the path integral framework
only if one adds to the classical action an $\hbar$ dependent term
that depends on the scalar curvature of the Riemannian metric
\cite{bdw}.

From the path integral perspective, these differences are directly
responsible for the absence of singularities in LQC, i.e., for the
resolution of the apparent tension we began our discussion with in
Sec.\ref{s1}. In the first form of the path integral, the
classically singular paths are not contained in the range of
integration since the paths in $b$ are bounded above while in the
classical singularity $b=\infty$. Furthermore the action is no
longer the classical action. In the second form of the path
integral, the domain of integration does include singular paths. But
now the fact that the action is corrected by quantum geometry
effects becomes crucial. Indeed, in this case the equations of
motion can be obtained explicitly by varying the action. These
equations and their solutions describe bouncing cosmologies which
are characteristic of the singularity resolution in LQC. Thus the
exact results on singularity resolution in LQC are in complete
harmony with the path integral intuition, once one realizes that the
action that descends from the Hamiltonian theory includes quantum
geometry corrections. Furthermore, because we have an additional
constant in the theory ---the Barbero-Immirzi parameter--- it is
meaningful to consider $\hbar$ expansions while retaining quantum
geometry effects. This is achieved by sharpening the precise manner
in which the limit is taken: $\hbar \to 0,\, \gamma \to \infty$ such
that $\hbar\gamma^3 = {\rm const}$. This $\hbar$ expansion enables
us to introduce the WKB approximation which helps one understand
features of the exact amplitude, e.g., the oscillatory versus
damping behavior that occurs as one varies the final configuration
$(\v_f, \p_f)$ keeping the initial configuration $(\v_i,\p_i)$
fixed. It also provides an `explanation' of the surprising
effectiveness of the effective equations \cite{vt} in LQC from a
path integral framework.

Thus, from the LQC perspective, it would be incorrect to simply
define the theory starting with smooth metrics and matter fields and
assigning to each path the weight that comes from the Einstein
Hilbert action because this procedure completely ignores the quantum
nature of the underlying Riemannian geometry. For a satisfactory
treatment of ultraviolet issues such as the singularity resolution,
it is crucial that the calculation retains appropriate memory of
this quantum nature. This viewpoint can be traced back to full LQG
and spinfoam models. In full LQG, quantum geometry is an essential
feature already of \emph{kinematics}. It is then not surprising that
in spinfoams the histories that one sums over are \emph{quantum}
geometries (captured in appropriate, colored 2-complexes). The
weight that is assigned to each history is motivated by the
Einstein-Hilbert action but does not descend \emph{directly} from
it, e.g. via a discretization procedure. Rather, one begins with a
the action of a constrained BF theory which is classically
equivalent with the Einstein Hilbert action but then incorporates
the (simplicity) constraint using considerations from quantum
geometry (quantum tetrahedron, representation theory and
interpretation of the Casimirs as eigenvalues of geometric
operators). Thus, the situation is parallel to the first form of the
path integral we obtained in this paper. In LQC we were fortunate in
that the integral over quantum paths could be recast into an
integral over all paths in the classical phase space. This enabled
us to carry out the steepest descent approximation and develop
physical intuition for the qualitative properties of the exact
extraction amplitude. A similar reformulation of spinfoams of the
full theory appears to be difficult. But if it could somehow be
achieved, one would have a powerful tool both to probe
semi-classical aspects of full quantum gravity and to develop
valuable intuition for the ultraviolet properties of the theory. In
particular, the resulting quantum geometry corrections to the full
Einstein-Hilbert action would bring the difference between spin
foams and perturbative path integrals into sharp focus.

\section*{Acknowledgments}

We thank Slava Mukhanov for raising the issue of the relation
between LQC and path integrals discussed in this paper. This work was
supported in part by NSF grants, PHY0854743, The George A.\ and
Margaret M.~Downsbrough Endowment and the Eberly research funds of
Penn State.

\appendix

\section{WKB approximation for constrained systems}
\label{app1}

Let us consider a system with phase space $\mathbb{R}^{2n}$ and
a single constraint $C(q,p) =0$, to be thought of as the
Hamiltonian constraint. We will assume that the $C(q,p)$ can be
written as a Taylor expansion in the $p$, as is the case for a
large class of physically interesting systems.

The kinematic Hilbert space is $L^2(\R^n)$ of normalizable wave
functions $\Psi(q) \equiv \Psi(q^1,\ldots, q^n)$. The elementary
operators are the usual $\hat{q}^j$ and $\hat{p_j}$ which act by
multiplication and derivation respectively:
\ba
\hat{q}^j \Psi(q) & = & q^j \Psi(q) \\
\hat{p}_j \Psi(q) & = & -i \hbar \frac{\partial \Psi}{\partial q^j}(q)  .
\ea
The physical states are then solutions to the quantum
constraint:
\be \label{consteq} \wh{C}\, \Psi = 0, \ee
where $\wh{C}$ is an operator analog of $C(q,p)$, obtained by
replacing $q,p$ with $\hat{q}^j$ and $\hat{p_j}$ with a suitable
choice of factor ordering. As is standard in the group averaging
procedure, we will assume that $\hat{C}$ is self-adjoint. For
unconstrained systems, the WKB ansatz provides approximate solutions
to the Schr\"odinger equation. In this Appendix we will extend that
method to obtain approximate solutions of (\ref{consteq}) where,
again, $\hbar$ plays the role of small parameter governing the
expansion. As one might expect, the main idea is to write both
$\wh{C}$ and $\Psi$ in (\ref{consteq}) as expansions in $\hbar$, and
to collect terms having the same $\hbar$ power. Let us now
explicitly calculate the zeroth and first order terms.

The construction is as follows. First, the constraint operator
$\wh{C}$ is written as a sum of `normal ordered' operators, in
which all $\hat{q}$'s appear to the left of the  $\hat{p}$'s:
\be \label{normord} \wh{C}= \sum_{n=0}^\infty  (\hbar/i)^n
C_n\big(\stackrel{\text{{\tiny
L}}}{\hat{q}},\,\stackrel{\text{{\tiny R}}}{\hat{p}}\big) . \ee
Here the $C_n$'s are functions on the classical phase space ---for
instance, $C_0$ will typically be the classical constraint
function--- which are now `evaluated' on the operators $\hat{q}$ and
$\hat{p}$ according to the `normal ordered' prescription indicated
by the superscripts. Second, the unknown state $\Psi(q)$ is written
as the exponential
\be \Psi(q)= \e^{\frac{i}{\hbar} S(q)} \ee
where the exponent is written as a power series in $\hbar$:
\be S(q) = \sum_{n=0}^\infty (\hbar/i)^n S_n(q). \ee

Since $C(q,p)$ is assumed to admit a Taylor expansion in the $p$, so
do $C_n(q,p)$. Imposition of the quantum constraint (\ref{consteq})
now leads to the following zeroth and first order equations:
\be
 C_0(q,\partial_q S_0) = 0,  \label{hjeq} \ee
\be \frac{1}{2}\left.\frac{\partial^2 C_0}{\partial p_i
\partial p_j}\right|_{p=\partial_q S_0}\frac{\partial^2
S_0}{\partial q^i \partial q^j} +\left.\frac{\partial
C_0}{\partial p_i}\right|_{p=\partial_q S_0} \frac{\partial
S_1}{\partial q^i}+C_1(q,\partial_q S_0) =0\label{conteq1} \ee
The zeroth order equation (\ref{hjeq}) can be recognized as the
Hamilton-Jacobi equation. The first order one, (\ref{conteq1}),
can be rewritten as follows. If we use the fact that $\wh{C}$
is self-adjoint, the condition $\wh{C}^\dag=\wh{C}$, when
applied to (\ref{normord}) implies \be \label{c1} C_1=
\frac{1}{2} \frac{\partial^2 C_0}{\partial q^j \partial p_j}.
\ee Using (\ref{c1}), and writing $a(q):= \e^{S_1(q)}$, Eq.
(\ref{conteq1}) can be written as a derivative of the function
$a$ along the vector field \be X := \left. \frac{\partial
C_0}{\partial p_j}\right|_{p=\partial_q S_0}
\frac{\partial}{\partial q^j} \ee as \be \label{conteq2} X(a)+
\frac{1}{2} a \, \text{div} X = 0. \ee

The divergence term in (\ref{conteq2}) suggest one to interpret
$a$ and $\Psi \sim a \, \e^{\frac{i}{\hbar}S_0}$ as half
densities on $\mathbb{R}^n$. Then (\ref{conteq2}) is just the
Lie derivative of $a$ along $X$:
\be \label{conteq3} \text{Eq. (\ref{conteq1}) } \iff
\mathcal{L}_X a = 0. \ee

These are the equations used in Sec. \ref{s4.2}.

\section{Exact Amplitude}
 \label{app2}

In Sec.\ref{s4.3} we compared the exact extraction amplitude
with the WKB approximation. In this Appendix we recall from
\cite{ach2} the expression that was used in the numerical
evaluation of the exact amplitude.

The first step in the calculation is to find the eigenvectors of
$\Theta$. They are given by $| k \pm \ket $, with $k>0$, satisfying
the eigenvalue equation
\be \Theta | k \pm \ket = 12 \pi G k^2\hbar^2 | k \pm \ket. \ee
These vectors are not normalized; the decomposition of the
identity reads
\be \label{identity}\mathbf{I}=\sint_{0}^{\infty}\frac{\dd k}{2
\pi k \sinh(\pi k)}\,\,\left(|k +\ket \bra k +|\,+\,|k -\ket
\bra k -|\right). \ee
In terms of the `volume basis' $|4 n \lo\hbar \ket $ used in
the main body of this paper, the vectors $|k \pm \ket $ are
given by
\be  \bra 4 n \lo\hbar | k \pm \ket = \left\{ \begin{array}{ll}
\sqrt{4 | n | \pi }i k P_{\pm n}(k) & \quad \pm n \geq 0 \\
0 & \quad \pm n <0 \end{array} \right. , \ee
where $P_{n}(k)$ is the following $(2n-1)$-degree polynomial in $k$:
\be P_{n}(k):= \frac{1}{i k (2n)!} \left. \frac{d^{2n}}{d
s^{2n}} \right|_{s=0} \left(\frac{1-s}{1+s} \right)^{i k} =
\sum_{m=0}^{2n} \frac{1}{m! (2n-m)!} \prod_{l=1}^{2n-1}(i k
+m-l) . \ee

We are now ready to present the expression of the extraction
amplitude. For this, it is convenient to work in the
deparameterized framework. In \cite{ach2} it was shown that the
extraction amplitude in the timeless framework coincides with
the transition amplitude of the deparameterized theory:
\be \label{exactamp} \G(\v_f,\p_f; \v_i, \p_i) = \bra \v_f |\,
e^{\f{i}{\hbar}\, \sqrt{\Theta}\, (\p_f-\p_i)}\, | \v_i \ket.
\ee
Let us take $\v_i = 4 \lo\hbar n_i$ and $\v_f = 4 \lo\hbar n_f$ with
$n_i$ and $n_f$ positive integers, and define $t:=\sqrt{12 \pi
G}(\p_f-\p_i)$.

By inserting the complete basis (\ref{identity}) in the right hand
side of (\ref{exactamp}), we obtain
\ba \G(\v_f,\p_f; \v_i, \p_i) & = &  -2 \sqrt{n_i n_f}\,
\sint_{0}^{\infty} \f{\dd k}{\sinh(\pi k)} \, P_{n_f}(k)\,
P_{n_i}(k)\, k e^{i k t} \\
& = &   -2 \sqrt{n_i n_f}\, P_{n_f}(-i \partial_t)\, P_{n_i}
(-i \partial_t)\, \sint_{0}^{\infty} \f{\dd k}{\sinh(\pi k)} \,
{k e^{i k t}} \\
& = & - \frac{\sqrt{n_i n_f}}{\pi^2}\, P_{n_f} (-i
\partial_t)\, P_{n_i}(-i \partial_t)\,
\psi^{(1)}(1/2-i\frac{t}{2 \pi}) \label{me} \ea
where $\psi^{(1)}(z)=d \log \Gamma(z)/d z$, and
$\Gamma(z)=(z-1)!$ is the Gamma function.

This last expression (\ref{me}) was the one used to numerically
compute the exact extraction amplitude for the plots in Sec.
\ref{s4.3}.

\end{document}